\documentclass[sigconf,nonacm,9pt]{acmart}

\newcommand\vldbdoi{XX.XX/XXX.XX}

\newcommand\vldbavailabilityurl{http://vldb.org/pvldb/format_vol14.html}

\usepackage[english]{babel}
\usepackage{caption}
\usepackage{subcaption}
\usepackage{algpseudocode}
\usepackage{lscape}
\usepackage[linesnumbered,ruled,vlined]{algorithm2e}
\usepackage{amsmath}
\usepackage{graphicx}
\usepackage{cleveref}
\usepackage{enumitem}
\usepackage{mathtools}

\newcounter{prob}
\newtheorem{problem}[prob]{Problem}
\newtheorem*{problem*}{Problem}

\newtheorem{ex}{Example}
\newtheorem{example}[ex]{Example}

\usepackage{dsfont}

\DeclareMathOperator*{\argmax}{argmax}

\def\HS{\hspace{\fontdimen2\font}}
\definecolor{darkgreen}{rgb}{0.15,0.55,0.15}
\definecolor{darkblue}{rgb}{0.1,0.1,0.5}
\definecolor{blue}{rgb}{0.01,0.40,.8}
\definecolor{darkgreen}{rgb}{0.15,0.55,0.15}
\definecolor{lightgreen}{rgb}{0.56,0.93,0.56}
\definecolor{lightred}{rgb}{1.0,0.58,0.58}
\definecolor{mred}{rgb}{.80,.12,.30}
\definecolor{grey}{rgb}{0.5,0.5,0.5}
\definecolor{Purple}{rgb}{.75,0,.85}
\definecolor{light-gray}{gray}{0.95}
\definecolor{mid-gray}{gray}{0.85}
\definecolor{darkred}{rgb}{0.7,0.25,0.25}
\definecolor{rose}{rgb}{1.0, 0.01, 0.24}
\newcommand{\red}[1]{\textcolor{red}{#1}}

\newcommand{\blue}[1]{\textcolor{blue}{#1}}
\newcommand{\indep}{\perp \!\!\! \perp}

\newcommand{\eat}[1]{}

\newcommand{\stitle}[1]{\vspace{2pt}\noindent\textbf{#1}}

\newcommand{\fdp}[0]{\texttt{FPM}\xspace}
\newcommand{\sys}[0]{\texttt{Mileena}\xspace}

\newenvironment{myitemize}{%
\begin{itemize}[leftmargin=1em, itemsep=.1em, parsep=.1em, topsep=.1em,
    partopsep=.1em]}
{\end{itemize}}

\author{Zezhou Huang}
\email{zh2408@columbia.edu}
\affiliation{
  \institution{Columbia University}
}

\author{Jiaxiang Liu}
\email{jl6235@columbia.edu}
\affiliation{
  \institution{Columbia University}
}

\author{Haonan Wang}
\email{hw2983@columbia.edu}
\affiliation{
  \institution{Columbia University}
}

\author{Eugene Wu}
\email{ewu@cs.columbia.edu}
\affiliation{
  \institution{DSI, Columbia University}
}

\begin{document}

\title{The Fast and the Private: Task-based Dataset Search}

\begin{abstract}
Modern dataset search platforms employ ML task-based utility metrics instead of relying on metadata-based keywords to comb through extensive dataset repositories. In this setup, requesters provide an initial dataset, and the platform identifies complementary datasets to augment (join or union) the requester's dataset such that the ML model (e.g., linear regression) performance is improved most.
Although effective, current task-based data searches are stymied by (1) high latency which deters users, (2) privacy concerns for regulatory standards, and (3) low data quality which provides low utility. 
We introduce \sys, a fast, private, and high-quality task-based dataset search platform. At its heart, \sys is built on pre-computed semi-ring sketches for efficient ML training and evaluation.  Based on semi-ring, we develop a novel {\it Factorized Privacy Mechanism} that makes the search differentially private and scales to arbitrary corpus sizes and numbers of requests without major quality
degradation.
We also demonstrate the early promise in using LLM-based agents for automatic data transformation and applying semi-rings to support causal discovery and treatment effect estimation.

\end{abstract}

\maketitle



\vspace*{-2mm}
\section{Introduction}

\begin{figure}
  \centering
      \includegraphics [width=0.4\textwidth]  
      {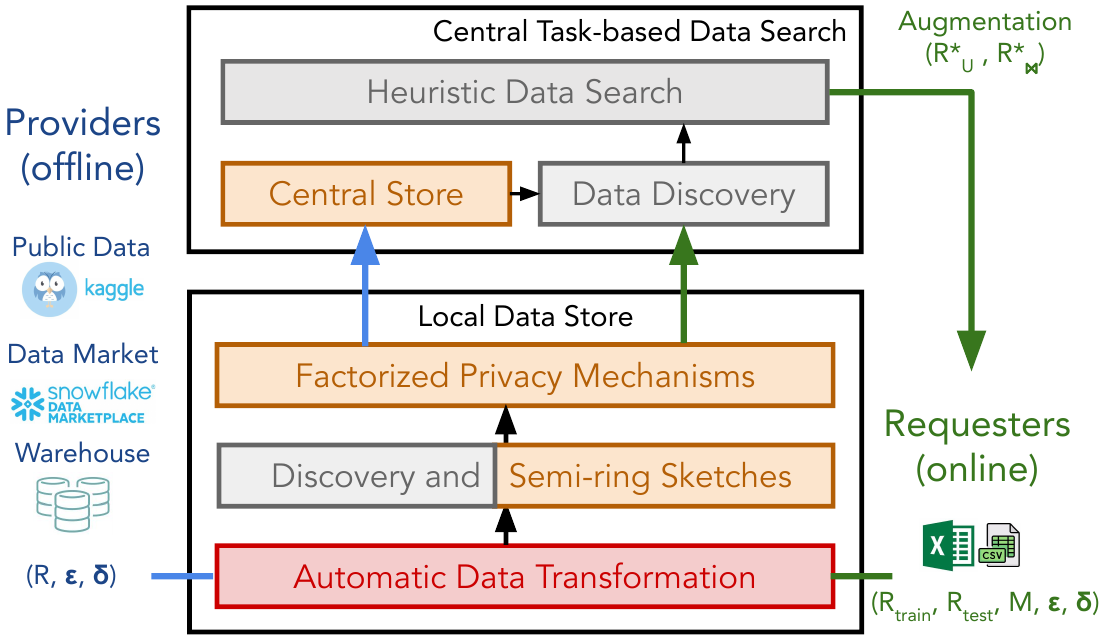}
      \vspace*{-4mm}
  \caption{\sys Architecture. Gray components are from previous works, orange represents current progress, and red indicates ongoing work. The blue workflow is offline for providers, while the green workflow is online for requesters.}
  \vspace*{-6mm}
  \label{fig:arch}
\end{figure}

Existing relational data repositories~\cite{nycopen,miloslavskaya2016big} offer the potential to augment and improve data-oriented tasks such as machine learning, and have motivated considerable work in both research~\cite{fernandez2018aurum,chepurko2020arda,galhotra2023metam} and industry~\cite{snowflakedatamarketplace,awsmarketplace,datamarketsurvey2022}.
The traditional approach uses keyword search~\cite{googledatasetsearch} over metadata about datasets but requires the user to guess relevant keywords, manually integrate each returned dataset, and assess its utility.  
 In response, recent work advocates for task-based data search~\cite{santos2022sketch,chepurko2020arda,nargesian2022responsible,li2021data,kitana}, which takes as input an ML task (based on training and test datasets) and returns datasets in the data store that augment the training dataset in a way to improve the model quality.  These augmentations can be any combination of joins and unions with other relations in the data store to add features, and  with relations to add more samples.   

In principle, such a system could enable a development cycle: as data users develop predictive or causal models over their local data, it searches for and automatically suggests or even integrates datasets that {\it concretely improves the user's model}.  
Existing approaches develop a data discovery index~\cite{fernandez2018aurum} to identify union and join candidate augmentations; but they laboriously evaluate each candidate by applying the augmentation, retraining, and cross-validating the model to assess {\it utility}.   Other works cluster and prune the candidates~\cite{galhotra2023metam}, but a single search query still takes minutes.   

We believe practical task-based data search remains stymied by three practical considerations:
\begin{myitemize}
\item {\it Latency}: 
Machine learning is often performed as part of an iterative user-facing data analysis and development process~\cite{psallidas2022data}.  As in web search, latency affects a user's willingness to use the system~\cite{arapakis2014impact}.  Unfortunately, existing data search systems take tens of minutes to hours because candidate assessment relies on costly model retraining and evaluation~\cite{chepurko2020arda,galhotra2023metam}.
These latencies pale in comparison to existing keyword-based dataset search services which, despite their disconnect from the user's data task, return results in seconds and dominate current deployments~\cite{snowflakedatamarketplace,googledatasetsearch}.

\item {\it Privacy}: Even within a single organization, data privacy, and access controls are major concerns.  For instance, access and use of data containing Personally Identifiable Information (PII) are regulated by governments~\cite{CCPA}.  Differential privacy is a promising approach that is rapidly gaining adoption due to its well-defined mathematical guarantees~\cite{dwork2006calibrating}.  It introduces noise to released statistics in a way that masks the presence or absence of an individual in the dataset, and was famously used to release fine-grained statistics for the 2020 US Census~\cite{kenny2021use}.
Unfortunately, existing applications of differential privacy have largely focused on federated machine learning use cases to train a single model.  When applied to dataset search over even a handful of datasets, it requires so much noise that search is no better than random.

\item {\it Data Quality}: 
It is now a common refrain that 80-90\% of effort in data science goes into data preparation and cleaning, and this is similarly true for machine learning applications~\cite{psallidas2022data}.
However, this challenge is exacerbated in a dataset search context, where data providers may upload hundreds or thousands of datasets and cannot be expected to prepare and clean each one.  It's similarly unrealistic to expect the end-user to do the same for each candidate augmentation.    Yet, preparation and cleaning may be necessary for search queries to return high-quality results that most improve the user's ML models.   To this end, a scalable, extensible, and {\it fully-automated cleaning procedure} is necessary.
\end{myitemize}

This paper describes our current progress to develop a fast, private, and high-quality task-based dataset search platform called \sys.  
At its heart, the system is built on the concept of semi-ring aggregation. 
Similar to data cubes which rely on aggregations that distribute across unions to pre-compute partial aggregates over partitions of a relation, semi-ring aggregates distribute across unions {\it and} joins.    Semi-rings have been designed for common statistical aggregation functions, as well as a wide range of machine learning models, including linear regression.  This is a natural match with dataset search, where the goal is to join and union an initial training dataset with registered relations, and evaluate a data task over the result.   For data tasks that can be formulated over semi-rings, such as training a linear model, the semi-ring computation can be pushed to the base relations and pre-computed.  

This insight helps \sys scale to thousands of datasets and return high-quality augmentations within a few seconds.   Data providers pre-compute and upload semi-ring aggregates of each dataset to \sys.   When a user submits a search request,  \sys uses a semi-ring-based proxy model to find the most promising augmentations, and then trains a final model that is returned to the user.    We also show that this approach is compatible with differential privacy, and develop a novel {\it Factorized Privacy Mechanism} that makes the entire search process differentially private, while scaling to arbitrary corpus sizes and numbers of requests without major degradation in the search result quality.

We also present our ongoing work that improves data quality and extends the semi-ring framework to causal inference.
For data quality, we propose an agent-based framework to automatically transform and extract features from a provider's dataset before registration with \sys.  The key idea is to use agents that perform a range of exploratory data analysis and context-gathering tasks to distill the semantics of the dataset into a compact representation, which is then used to generate transformation and featurization functions custom to the dataset.      
To support causal inference, we are studying how semi-rings can be used for both causal discovery and estimating average treatment effects, and novel challenges that arise in the context of dataset search.    By building on semi-rings, causal inference tasks are automatically made differentially private.

\vspace*{-2mm}
\section{Problem and Solution}

In this section, we define the problem of differentially private task-based dataset searches. 
We start with the background of data model, ML task, and differential privacy.
Then, we lay out the trust model, privacy needs, and search problem. 
Finally, we walk through the architecture of \sys to solve this problem.
While non-private search has been studied~\cite{santos2022sketch,chepurko2020arda,nargesian2022responsible,li2021data}, we are the first to establish the trust and privacy requirements for practical dataset search.

\vspace*{-2mm}
\subsection{Problem Definition}
\stitle{Data Model.}  
We follow the standard relational data model.  Relations are denoted as $R$, attributes as $A$, and domains as $dom(A)$. 
For clarity, the schema is included in square brackets $R[A_1,\cdots,A_n]$.

\stitle{Data Task.}  We focus on  ML task $(M,R_{train},R_{test})$, which seeks to train a good model on a relation $R$ containing features $X\subset S_R$ and target $Y \in S_R$.  The function $M.Train(R_{train})$ returns a  model $m$ that predicts $y$ from $X$. The function $M.Evaluate(m, R_{test})$ returns the {\it task utility} (typically cross-validation accuracy) on a test dataset. The goal of dataset search is to augment $R_{train}$ with additional features (via joins) and samples (via unions) so that the {\it utility} over the augmented dataset is maximized.  
Beyond ML tasks, \Cref{sec:ci} describes our ongoing work to support causal inference (CI) tasks.

\stitle{Trust Model.}  
For dataset search,
the $1^{st}$-level aggregators (providers/ requesters)  share data with the $2^{nd}$-level aggregator (central search).
However, the data might contain Personally Identifiable Information (PII). To comply with legal standards~\cite{CCPA}, organizations must protect data against misuse by untrusted entities.

\begin{figure}
  \centering
      \includegraphics [width=0.35\textwidth]  {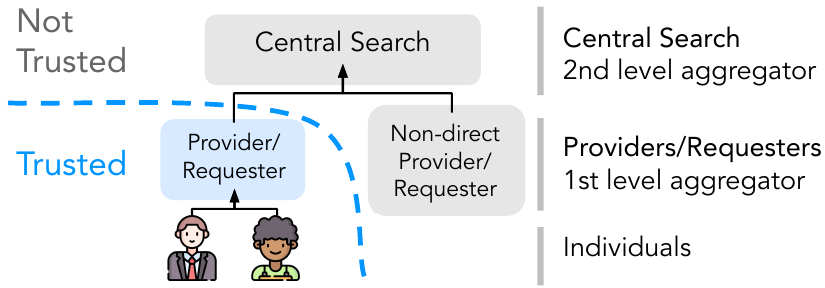}
      \vspace*{-3mm}
  \caption{\sys trust model:  individuals only trust the  direct $1^{st}$-level aggregator, and not any others. }
  \label{fig:trust_model}
  \vspace*{-5mm}
\end{figure}

To model trust, on the one extreme, the local trust model used by Apple~\cite{cormode2018privacy} and Google~\cite{erlingsson2014rappor} assumes that individuals don't trust any aggregator. While this requires weak assumption, its mechanisms provide limited utility~\cite{wei2020federated}. 
At the other extreme, others assume a global trust model ~\cite{johnson2018towards}, where individuals trust the central aggregator (central search). 
Mechanisms for this model yield high utility but the assumption is unrealistic. 
For dataset search, we introduce a two-tiered trust model (\Cref{fig:trust_model})  inspired by private federated ML~\cite{wang2020hybrid}.  
Here,  individuals trust direct $1^{st}$-level aggregators (e.g., patients trust direct hospitals) but do not trust $2^{nd}$-level aggregator and other non-direct $1^{st}$-level aggregators (e.g., other hospitals).

\stitle{Differential Privacy.} For untrusted entities, rather than prohibit access outright, differential privacy (DP)~\cite{dwork2006calibrating} supports analysis of sensitive data while bounding the degree of privacy loss based on the budget $(\epsilon,\delta)$ set by each aggregator.  Each query on the dataset adds noise to the results, inversely proportional to the budget consumed; when $\epsilon=0$,
the dataset becomes inaccessible.
Formally:
\vspace*{-2mm}
\begin{definition}[$(\epsilon,\delta)$-DP]
Let $f$ be a randomized algorithm that takes a relation $R$ as input. $f$ is $(\epsilon, \delta)$-DP if, for all relations $R_1, R_2$ that differ by adding or removing a row, and for every set $S$ of outputs from $f$, 
$Pr[f(R_1) \in S] \leq e^{\epsilon} Pr[f(R_2) \in S] + \delta$,
where $\epsilon$ and $\delta$ are non-negative real numbers (called privacy budget). $\epsilon$ controls the level of privacy, and $\delta$ controls the level of approximation.
\end{definition}

\stitle{Differentially Private Task-based dataset search.} Given a data corpus with datasets from different providers, a requester sends a request with datasets to augment a task (e.g., ML). The goal is to identify a set of augmentable (join/union) datasets that maximize task utility while satisfying trust and privacy requirements.

To formalize this, let  $\mathcal{R} = \{R_1, R_2,...\}$  be a data corpus collected from some providers.
Providers and requesters hope to disclose datasets to the untrusted central search. Each sets a \texttt{DP} budget $(\epsilon_i, \delta_i)$ for each dataset $R_i$, which is independent of other datasets and the central search. Requester sends a request with training and testing dataset $(R_{train},R_{test})$, chooses a model $M$, and specifies $(\epsilon, \delta)$. Requester's goal is to train model $M$ on $R_{train}$ and maximize its performance on $R_{test}$, which we call the task's {\it utility}.

To maximize the {\it utility}, the requester aims to find a set of provider datasets in $\mathcal{R}$ to augment data. The function $Discover(R, augType)$ finds datasets in $\mathcal{R}$ that can be joined or unioned with $R$, given $augType \in \{\Join, \cup\}$. 
Putting everything together:

\noindent\begin{problem}[Task-Based dataset search.]
\label{searchprob}
For each request $(R_{train},R_{test}, M, \epsilon, \delta)$, find the set of datasets $\mathbf{R}^*_\cup, \mathbf{R}^*_\Join  \subseteq \mathcal{R}$ such that
\begin{flalign*}
\mathbf{R}^*_\cup, \mathbf{R}^*_\Join=&\quad\argmax_{\mathbf{R}_\cup, \mathbf{R}_\Join}   M.Evaluate(m, R_{testAug})\\
s.t. &\quad \mathbf{R}_\cup \subseteq Discover(R, \cup), \mathbf{R}_\Join \subseteq Discover(R, \Join),\\
&\quad R_{trainAug} = (R_{train} \cup_{R_1\in\mathbf{R}_\cup}  R_1)\Join_{R_2\in\mathbf{R}_\Join} R_2\\
&\quad R_{testAug} = R_{test} \Join_{R\in\mathbf{R}_\Join} R\\
&\quad m = M.Train(R_{trainAug})\\
&\quad \text{The search over $(R_{train},R_{test})$ is  $(\epsilon, \delta)-DP$}\\
&\quad  \text{The search over $R_i$ is $(\epsilon_i, \delta_i)-DP$, $\forall R_i\in\mathcal{R}$}
\end{flalign*}
\end{problem}

\subsection{\sys Walkthrough}

In this section, we walk through \sys's architecture (\Cref{fig:arch}) to solve \Cref{searchprob}.
\sys stands on prior  data discovery and search systems~\cite{fernandez2018aurum, chepurko2020arda,castelo2021auctus}. However, these systems rely on slow join/union operations and fail to meet privacy requirements. To improve upon this, we use (1) a proxy model that can quickly estimate the benefits of a candidate augmentation using pre-computed semi-ring sketches (\Cref{sec:fac_aug}), (2) a factorized privacy mechanism to ensure DP (\Cref{s:private_search}), and (3) an agent-based automated data transformation framework to improve data quality (\Cref{sec:transform}).

\subsubsection{Local Data Store.}
This locally manages each provider's/requester's raw data.  It transforms and pre-processes each dataset, and generates privatized sketches that will be uploaded to the centralized search platform.   Providers specify a DP budget for each relation they register; requesters upload training (and possibly testing) datasets with their DP budgets, and a task $M$.

\stitle{Automatic Data Transformation.} 
Raw datasets are noisy and require parsing and transformations to derive predictive features. 
We propose a fully automated approach based on LLM agents (\Cref{sec:transform}), so the transformed dataset is most useful to search tasks.    These costs can scale to the provider and requester's willingness to bear them.  An enterprise may allocate resources to datasets in a data lake to improve searchability, while data sellers already clean and prepare data that they provide in existing data markets~\cite{snowflakesummit}.

\stitle{Discovery and Semi-ring Sketches.} 
Data discovery indexes datasets based on their schemas and column features to quickly find join and union candidates (that need to be further assessed against the data task).   We currently use min-hash and TF-IDF sketches based on Aurum~\cite{fernandez2018aurum} to search for augmentation datasets based on column similarity.
For utility assessment, we propose
novel semi-ring sketches, which are used for efficient ML training and evaluation\footnote{Semi-aggregates can express complex models like linear regression~\cite{schleich2016learning}, gradient boosting and random forests~\cite{joinboost}, and approximate generalized linear models~\cite{huggins2017pass}.}.

\stitle{Factorized Privacy Mechanism.} 
Semi-ring sketches are aggregates over raw data, and can be privatized by adding appropriate noise. Compared to standard privacy mechanisms for data discovery index~\cite{fernandes2021locality, weggenmann2018syntf}, we introduce a novel Factorized Privacy Mechanism for semi-ring sketches (\Cref{s:private_search}). Once privatized, these sketches can be repeatedly used across searches without any privacy cost.

\subsubsection{Central Task-based dataset search.} 
\label{sec:central}
For each request, the central task-based dataset search component solves \Cref{searchprob} online.

\stitle{Central data store.}   By default, all provider and requester data is privatized before upload to the central data store, and all search algorithms are over these privatized sketches.

\stitle{Data Discovery.} For each request, we employ Aurum~\cite{fernandez2018aurum} to discover augmentable data; Aurum uses pre-computed sketches to retrieve augmentable data based on the column Jaccard similarity  (minhash sketches) and cosine similarity (TF-IDF sketches).

\stitle{Search algorithm.} Given the candidate augmentations, the search algorithm greedily finds a sequence of vertical and horizontal augmentations that maximizes expected task utility.  The basic algorithm iterates over each augmentation, materializes the augmented dataset, trains the model, evaluates its training accuracy (or other quality measure), selects the augmentation that most improves the utility, and repeats.   Training and evaluation are so expensive that existing works~\cite{chepurko2020arda, castelo2021auctus,galhotra2023metam} primarily focus on aggressively pruning the set of augmentation to evaluate. 
In addition to these pruning methods, we use a semi-ring-compatible proxy model (e.g., linear regression) to directly derive the augmented model parameters and compute the model's utility in time independent of the relation sizes.   This allows us to evaluate candidates in milliseconds.  When used to power an AutoML service, \sys  improves final model accuracy, and reduces both query latency and monetary cost by orders of magnitude as compared to existing dataset search platforms and AutoML services (\Cref{fac:exp}).

\section{Private Semi-ring Sketches}

In this section, we delve into Semi-ring Sketches, exploring how they can efficiently train and evaluate the proxy model (linear regression), as well as the associated differential privacy mechanism.

\subsection{Semi-ring Aggregates Primer} \label{s:backgroundmsgpassing}

Augmentations are composed of joins and unions ($\mathbf{R}^*_\cup, \mathbf{R}^*_\Join$ in \Cref{searchprob}) with relevant datasets. For efficient heuristic data search, it is necessary to reevaluate the data task (e.g., retraining and evaluating a ML model) after augmenting (join or union) the training data in time independent of the relation sizes.  Our main observation is that semi-ring aggregations are a natural fit for this use case.

\stitle{Annotated Relations and Semi-ring Aggregates.} The annotated relational~\cite{abo2016faq} model maps $t\in R$ to a commutative semi-ring $(D, +, \times, 0, 1)$, where $D$ is  a set, $+$ and $\times$ are commutative binary operators closed over $D$, and $0/1$ are zero/unit elements. An annotation for $t \in R$ is denoted as $R(t)$. Semi-ring annotation expresses various aggregations. E.g., the natural number expresses count.

\stitle{Semi-ring Aggregation Query.} Semi-ring aggregation queries can now be reformulated using annotated relations by translating group-by, union, and join operations into addition ($+$) and multiplication ($\times$) operations over the semi-ring annotations, respectively:
\begin{myitemize}
\item $(\gamma_\mathbf{A} R)(t) =  \sum \{R(t_1) | \HS t_ 1 \in R , t = \pi_{\mathbf{A}} (t_1 )\}$. The annotation for group-by $\gamma_\mathbf{A} R$ is the sum of the annotations within the group. 
\item  $(R_1\cup R_2)(t) = \HS R_1(t) + R_2(t)$. The annotation for union $R_1 \cup R_2$ is the sum of annotations in  $R_1$ and $R_2$.
\item $(R_1\Join R_2)(t) =\HS R_1(\pi_{S_{R_1}} (t)) \times R_2(\pi_{S_{R_2}} (t))$. The annotation for join $R_1 {\Join} R_2$ is the product of contributing annotations.
\end{myitemize}

\stitle{Aggregation Pushdown.}
Semi-rings let us distribute aggregations through joins and unions~\cite{abo2016faq}. Consider the query $\gamma_D (R_1[A,B] \Join R_2[B,C] \Join R_3[C,D])$. Rather than apply $\gamma$ after the join (which is $O(n^3)$ where $n$ is relation size), $\gamma$ can be performed on $R_1$ (and similarly on $R_2,R_3$) before joining with $R_2$, in $O(n)$:
$\gamma_{D} (\gamma_{C} (\gamma_{B} (R_1[A,B])\Join R_2[B,C]) \Join R_3[C,D])$.
Associativity of additions can be exploited for union: 
$\gamma_A (R_1[A,B] \cup R_2[A,B]) = \gamma_A (R_1[A,B]) \cup \gamma_A (R_2[A,B])$.

\stitle{Application to ML.} Such aggregation pushdown optimization has been applied to accelerate ML, by designing different semi-rings for different ML models. We use linear regression~\cite{schleich2016learning} as an example. Given the training data  $\mathbf{X} \in \mathbb{R}^{n \times m}$, and the target variable $\textbf{y} \in \mathbb{R}^{n\times 1}$, linear regression computes $\theta^*{=}(\mathbf{X}^T\mathbf{X})^{-1}\mathbf{X}^T\textbf{y}$. By considering $\textbf{y}$ as a special feature,  we find that $\mathbf{X}^T\mathbf{X}{\in}\mathbb{R}^{m \times m}$ is the core sufficient statistics to compute, where each cell represents the sum of products between feature pairs.
To compute $\mathbf{X}^T\mathbf{X}$ over the join, we design the covariance matrix semi-ring~\cite{schleich2016learning} as a triple $(c,\mathbf{s},\mathbf{Q})\in(\mathbb{Z},\mathbb{R}^{m}, \mathbb{R}^{m \times m})$, which contains the count, sums, and sums of pairwise products.  
$+$ and $\times$ are defined as:
$a + b = (c_a + c_b,\mathbf{s}_a + \mathbf{s}_b, \mathbf{Q}_a + \mathbf{Q}_b)$, $
a \times b = (c_a c_b,c_b\mathbf{s}_a + c_a\mathbf{s}_b, c_b\mathbf{Q}_a + c_a\mathbf{Q}_b + \mathbf{s}_a \mathbf{s}_b^T +\mathbf{s}_b \mathbf{s}_a^T)$.
Then, computing $\mathbf{X}^T\mathbf{X}$ is reduced to executing $\gamma(R_1 \Join ... \Join R_k)$, where aggregation can be pushed down as discussed before.

\begin{figure}
  \centering
      \includegraphics [width=0.4\textwidth]  {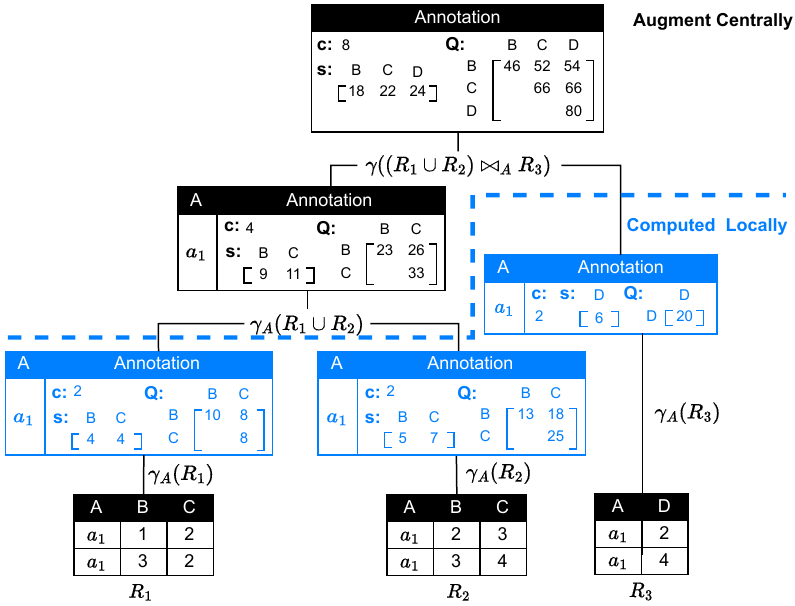}
      \vspace*{-3mm}
  \caption{$\gamma((R_1\cup R_2)\Join_A R_3)$ computes linear regression; \blue{aggregations} are pushed before joins, and are pre-computed locally to accelerate central data search. 
  }
  \vspace*{-5mm}
  \label{fig:factorizedmlexp}
\end{figure}

\begin{example}
\label{example:fac}
Consider  $R_1, R_2, R_3$ in \Cref{fig:factorizedmlexp}. We aim to train linear regression on $(R_1{\cup} R_2){\Join_A} R_3$ using D as the feature and C as the target variable.
The naive solution is to first materialize the union and join results and then compute $\mathbf{X}^T\mathbf{X}$. 
Using semi-ring, we can optimize the query plan (\Cref{fig:factorizedmlexp}) by pushing down aggregations: $\gamma\left( (\gamma_A(R_1){\cup}\gamma_A(R_2)){\Join_A\gamma_A(R_3)} \right)$. This approach yields the same result as the naive solution, but avoids the costly materialization.
\end{example}

\subsection{Pre-computed Semi-ring Sketches}
\label{sec:fac_aug}

Although aggregation pushdown optimizes training and evaluation for a single augmentation, the whole search process still requires the recomputation of semi-ring aggregates across all augmentations. To optimize this, we aggressively pre-compute aggregate as sketches locally. 
Online, the evaluation for horizontal augmentation reduces $O(n)$ to $O(1)$ and $O(d)$ for vertical augmentation, with $n$ being the relation size and $d$ the join key cardinality. Typically, $d<<n$.

\subsubsection{Horizontal Augmentation}
Horizontal augmentation $A^h$ unions training data $R$. The aggregation can be pushed before union:
$\gamma(R \cup A^h)
  = \blue{\gamma(R)} \cup  \red{\gamma(A^h)}$. The key optimization is to pre-compute $\blue{\gamma(R)}$ and 
$\red{\gamma(A^h)}$ when data is uploaded to local data store.
$\blue{\gamma(R)}$ is shared across all candidate horizontal augmentations, and $\red{\gamma(A^h)}$ is shared across user requests. Now, horizontal augmentation adds the pre-computed aggregates in near-constant time.

\subsubsection{Vertical Augmentation}
Vertical augmentation is more complex because pushing 
aggregation through the join needs to take the join key $j$ into account.  
Consider $A^v$, which joins the training data $R$  using join key $j$: $\gamma(R \Join_{j}  A^v) 
  = \gamma(\blue{\gamma_{j}(R)} \Join_{j} \red{\gamma_{j}(A^v)})$.
$\blue{\gamma_{j}(R)}$ is shared among all vertical augmentation candidates with join key $j$. Thus, we pre-compute  $\blue{\gamma_{j'}(R)}$ for all of its valid join keys $j'$. 
We also pre-compute $\red{\gamma_{j'}(A^v)}$ for all of its valid join keys, and share them across all requests where $A^v$ is a vertical candidate. 

\begin{figure}
    \centering
    \includegraphics[width=0.8\columnwidth]{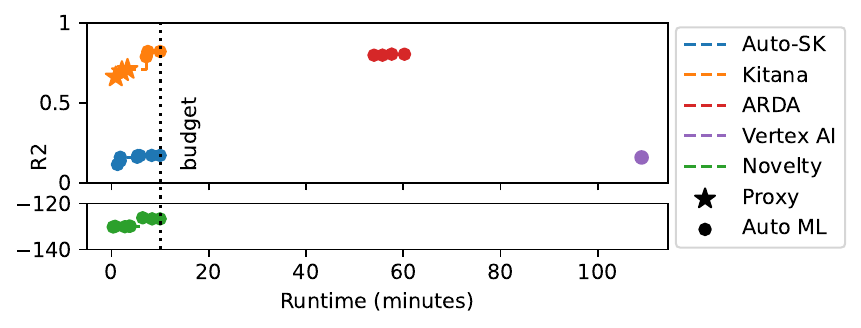}
    \vspace*{-5mm}
    \caption{Task utility (testing $R2$) with $10$ minutes time budget (dotted line). \sys searches a corpus of 517 datasets with linear regression and reaches $R2=\sim0.7$; \sys then sends the augmented dataset to AutoML and further improves $R2$ to $0.82$. Other baselines are either slow or have low $R2$.}
    \label{fig:moneyfig}
    \vspace*{-5mm}
\end{figure}

\subsubsection{Experiments.} 
\label{fac:exp}
We use \sys to power an AutoML service that uses up to 10 minutes for data search, then materializes the augmented dataset and uses the remaining time to run an AutoML library or service~\cite{kitana}.   We compared it with existing ML-based dataset search systems  (ARDA~\cite{chepurko2020arda}, Novelty~\cite{li2021data}), AutoML libraries (Auto-sklearn), and Google's Vertex AI on 517 datasets from NYC Open Data~\cite{nycopen}.
\Cref{fig:moneyfig} reports $R2$ from \sys's proxy model (stars) and the AutoML models (circle).   Note that ARDA and Vertex AI don't enforce the time budgets.
Pure AutoML approaches perform poorly because the dataset lacks predictive features; 
ARDA eventually finds a good model after ${\approx}50$min that is slightly worse than \sys.
Novelty assesses augmentations based on how ``novel'' the data is compared to the training data, but is uncorrelated with model utility and actually degrades the final model.   \sys returns a high quality model almost immediately, and converges to the highest quality model within the budget.

\subsection{Factorized Privacy Mechanism}
\label{s:private_search}
Private task-based data search is particularly challenging: even a single request requires model retraining across a vast number of augmented datasets as join/union results. How can we ensure a scalable data search with large number of datasets and requests, without depleting the privacy budgets of both requesters and providers?

Our key insight is that the semi-ring sketches, once privatized,
are composable (through semi-ring operators) and reusable (as post-processing without additional DP cost), making them ideal for data search that trains ML across joins and unions from different sets of relations. We name our mechanism {\it Factorized Privacy Mechanism} (\fdp), which applies Gaussian mechanism~\cite{dwork2006our} to these sketches (\blue{blue} in \Cref{fig:factorizedmlexp}) locally before transferring to the central data corpus.  We further develop novel budget allocations that optimize the proxy model's accuracy~\cite{saibot}.

\begin{figure}
  \centering
     \begin{subfigure}[t]{0.14\textwidth}
         \includegraphics[width=\textwidth]{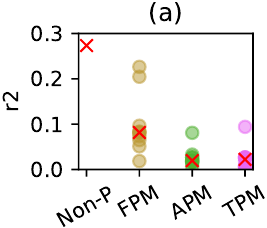}
         \vspace*{-5mm}
     \end{subfigure}
     \begin{subfigure}[t]{0.12\textwidth}
         \includegraphics[width=\textwidth]{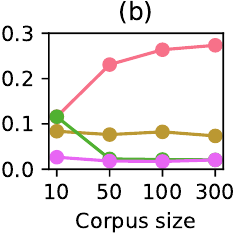}
         \vspace*{-5mm}
     \end{subfigure}
     \begin{subfigure}[t]{0.19\textwidth}
         \includegraphics[width=\textwidth]{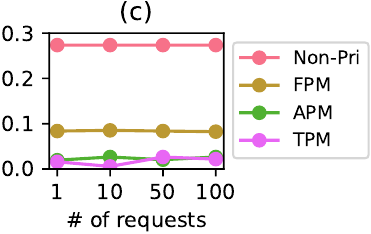}
         \vspace*{-5mm}
     \end{subfigure}
     \vspace*{-4mm}
  \caption{Task utility (non-private $r2$) for ML over augmented dataset from different private searches (a) across $10$ runs with the median as a red cross, (b) varying the corpus size, and (c) varying the number of requests. \fdp notably outperforms  even with a high volume of corpus datasets and requests. }
  \label{exp:nycutility}
  \vspace*{-6mm}
\end{figure}

\subsubsection{Experiments}
Using NYC Open Data and regression models (utility is $R2$), \Cref{exp:nycutility} shows how \fdp scales far beyond existing mechanisms in terms of corpus size and number of search requests as compared to 
\textbf{APM}~\citep{wang2018revisiting}, which applies DP mechanism to aggregates after computing the join/union results under a global trust model, 
and  \textbf{TPM}~\cite{yang2020local}, applies DP mechanism to individual tuples.
\textbf{\texttt{Non-P}} reports results without any privacy; although \fdp achieves up to ${\sim}40{-}90\%$ of the Non-P utility, the gap can be further reduced by clustering and smoothing join groups~\cite{kellaris2013practical} in the future works.

\vspace*{-2mm}
\section{Ongoing Work}

We now describe two directions of ongoing work.
\vspace*{-2mm}

\subsection{Hand-Free Data Transformation}
\label{sec:transform}

Data transformation is pivotal for ML~\cite{sambasivan2021everyone}, but \sys uses private sketches rather than transformable raw data during the search.  
Is it possible to transform datasets locally prior to sketch computation in ways that benefit a variety of data tasks?

Recent data transformation approaches rely on deep-learning~\cite{mei2021capturing,heidari2019holodetect}, including Language Learning Models (LLMs)~\cite{narayan2022can}.
Although powerful and promising, LLMs need to serialize datasets into a textual form to include in the prompt context.  In contrast, LLMs have limited context lengths (GPT-4 supports 8K input tokens) and are very costly (GPT-4 costs $\$0.03/1K$ tokens).  Long contexts also lead to unreliably attention and hallucinations~\cite{liu2023lost}.

Our intuition is that developers and data scientists do not design features in one-shot from the raw data either.  They instead perform exploratory data analysis (EDA), and understand the problem context---both to reduce the amount of information to keep in their human memory, and to identify the salient semantics relevant to the problem---and then use the knowledge to synthesize features that incorporate the problem semantics.   To this end, we propose an agent-based framework for data transformation.  Each agent specializes in a particular task, and summarizes the information in a form consumable by an LLM or another agent. The design is illustrated in \Cref{fig:agentarch}, with the following agents:

\begin{myitemize}

\item {\bf EDA}: This agent explores data and related docs to suggest transformations.
Our implementation inputs the ML task contexts, a sample of ten rows, and column aggregates (min, max, median), and let this agent outputs a list of data transformations in NL.

\item {\bf Coder}: Each suggested transformation by EDA is designated to one {\bf Coder}, which also inputs the related column samples and outputs a Python function to implement the transformation.

\item {\bf Debugger}:
This agent inputs the function, accesses a Python environment, and 
ensures that the function can run. Following ~\cite{sakib2023extending}, Debugger iteratively modifies the function based on error messages.
By default, the debugging is retried up to $10$ times; if it still fails, that transformation is ignored.

\item {\bf Reviewer}: 
This agent evaluates
the outputs from Debugger to ensure transformations meet EDA's requirements. 
It reviews the sample transformed data, and confirms if it aligns with the NL description by EDA to finalize the transformation.

\end{myitemize}

\subsubsection{Experiments}
\begin{figure}
  \centering
      \begin{subfigure}[t]{0.14\textwidth}
         \includegraphics[width=\textwidth]{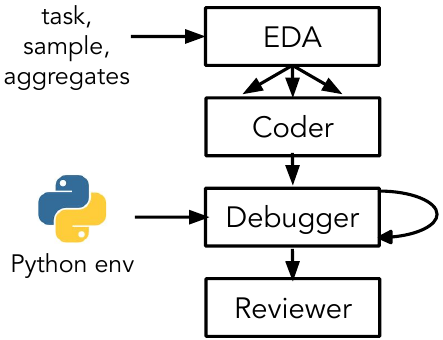}
         \vspace*{-5mm}
         \caption{}
         \label{fig:agentarch}
     \end{subfigure}
     \begin{subfigure}[t]{0.2\textwidth}
     \includegraphics[width=\textwidth]{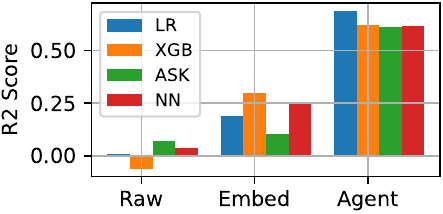}
         \vspace*{-5mm}
         \caption{}
         \label{fig:agentresult}
     \end{subfigure}
     \vspace*{-3mm}
  \caption{(a) Architecture for Agent-based Data Transformation. (b) Agent-based Transformation shows Model Performance (R2) across different ML models. }
  \label{fig:semantic_tranformation}
  \vspace*{-6mm}
\end{figure}

We evaluated linear regression, XGBoost, Auto-Sklearn, and TabNet~\cite{arik2021tabnet} (SOTA DNN for tabular data) on the Kaggle Airbnb data~\cite{airbnb}, and report model R2. We compare no transformations, transformations using GPT-4 agents (us), and transformations using ada-002 embeddings (which create high-dimensional features for string columns).
\Cref{fig:agentresult} shows that agent-based transformation trumps model complexity and transformation approaches. 
The agents suggested diverse and useful transformations, from standard one-hot-encoding to more complex ones, such as string extraction and calculating stay duration from date strings.
The most exciting result is that, \textbf{with agent-based transformations, linear regression (which is easy to maintain, and fast to train \& predict as compared to NN models) out-performed all others}.

Our ongoing work moves toward a data store that continuously evolves and improves the transformations.  This introduces novel challenges, such as how to efficiently update the sketches under DP~\cite{huang2021frequency}, incorporate new information about datasets over time (e.g., crawl the web for data documentation), manage a library of diverse agents, and interact with the central data store.

\vspace*{-2mm}
\subsection{Causal Inference}
\label{sec:ci}

Causal Inference (CI)
seeks to answer questions such as ``What is smoking's impact on cancer?"  It is distinguished from standard ML which focuses on learning correlations (e.g., ``how correlated is smoking to cancer?'') because relationships are asymmetric and the causal direction matters. CI queries rely on an accurate causal model, represented as a directed acyclic graph (DAG). Without key confounders~\cite{youngmann2023causal}, its intervention estimates can be arbitrarily incorrect.  Dataset search offers the potential to {\it find} these missing variables and {\it discover} a sufficient causal model.     

\stitle{Factorized Causal Discovery.} 
Existing causal discovery algorithms~\cite{glymour2019review} only infer an equivalence class of DAGs given observational data alone. For example, conditional independence tests may determine that smoking and cancer are dependent, but cannot infer smoking causes cancer. Fortunately, the causal direction can be determined assuming sufficiency (no unobserved confounders), non-Gaussian noise and linear relationships~\cite{shimizu2011directlingam}.  Consider, $X$ and $Y$ with $X \sim U(0, 10)$ and $Y = 2X + \epsilon \sim U(0, 10)$. A linear regressor using $X$ to predict $Y$ yields residuals $res_y \indep X$ because $\epsilon \indep X$, but the residuals when using $Y$ to predict $X$ yields $res_x \not\indep Y$.

However, the assumptions are not realistic in practice.   Fortunately, 1-N and N-N relationships between relations create colliders, on a lifted representation of variables, discoverable by conditional independence tests~\cite{maier2013sound} that relax the linearity and non-Gaussian noise assumptions.  Our ongoing work focuses on using semi-rings to integrate these ideas into \sys's fast and DP framework.

\stitle{Differentially Private Treatment Effects.} Given a causal diagram, existing techniques~\cite{salimi2020causal} that evaluate treatment effects require the distribution of treatment, target, and adjustment variables, potentially from different relations. This requires joining privatized relations, which may amplify DP noise so much that render the resultant join distribution ineffective. However, \cite{lee2020causal} shows that treatment effects are computable from marginal distributions.  

We ran a synthetic experiment with 3 relations $R_1(T,Y)$, $R_2(T,G)$, $R_3(P,A,Y)$; with binary attributes student qualification ($T$), overall score ($Y$), gender ($G$), student participation ($P$), and assignment completion ($A$), DP budgets $\epsilon = 1$ and $\delta = 10^{-6}$, and 1-to-1 relationships.  The causal diagram is $T \rightarrow P \rightarrow A \rightarrow Y$ and $D$ is a confounder between $T$ and $Y$ ($T\leftarrow D\rightarrow Y$). We intervene on $T$ and estimate its expected effect on $Y$, $E[Y \ | \ do(T = 1)]$ (ATE = $E[Y \ | \ do(T = 1)] - E[Y \ | \ do(T = 0)]$). We compare relative error using (1) backdoor adjustment by estimating $P(X, Y, G)$ from privatized $R_1$ and $R_2$, then $R_1 \Join R_2$. (2) $\sum_y y\sum_{a} P(a \ | \ t) \sum_{p} P(y \ | \ a, p)P(p)$ by estimating $P(A, T)$  from privatized $R_1$ and $R_3$, then $R_1 \Join R_3$ along with a noisy histogram of $R_3$.  Their respective relative errors were $10.25\%$ and $0.21\%$.  Surprisingly, splitting the privacy budget between $R_3$ and its histogram greatly improves estimate accuracy.
Our ongoing work designs intermediates from each dataset based on semi-ring aggregations to (1) represent the marginal distribution of the dataset, and (2) support computation of the joint distribution through joins.
\vspace*{-3mm}

\bibliographystyle{ACM-Reference-Format}
\bibliography{main}


\end{document}